\documentclass[aps,prl,twocolumn,showpacs,superscriptaddress,final]{revtex4-1}
\usepackage[final]{graphicx}
\usepackage{amsmath}
\usepackage{color}
\usepackage{epstopdf}
\usepackage{float}
\usepackage{textcomp}
\usepackage[hidelinks = true]{hyperref}

\begin{document}

% ***** TITLE
\title{Mode-selective coupling of coherent phonons to the Bi2212 electronic band structure}

% ***** AUTHORS
\author{S.-L. Yang}
\altaffiliation[current affiliations: ]{Kavli Institute at Cornell for Nanoscale Science; Laboratory of Atomic and Solid State Physics, Department of Physics; Department of Materials Science and Engineering. Cornell University, Ithaca, New York 14853, USA}
\affiliation{Stanford Institute for Materials and Energy Sciences, SLAC National Accelerator Laboratory, 2575 Sand Hill Road, Menlo Park, CA 94025, USA}
\affiliation{Geballe Laboratory for Advanced Materials, Departments of Physics and Applied Physics, Stanford University, Stanford, CA 94305, USA}
\author{J.~A. Sobota}
\affiliation{Stanford Institute for Materials and Energy Sciences, SLAC National Accelerator Laboratory, 2575 Sand Hill Road, Menlo Park, CA 94025, USA}
\affiliation{Advanced Light Source, Lawrence Berkeley National Laboratory, Berkeley, CA 94720, USA}
\author{Y. He}
\author{D. Leuenberger}
\affiliation{Stanford Institute for Materials and Energy Sciences, SLAC National Accelerator Laboratory, 2575 Sand Hill Road, Menlo Park, CA 94025, USA}
\affiliation{Geballe Laboratory for Advanced Materials, Departments of Physics and Applied Physics, Stanford University, Stanford, CA 94305, USA}
\author{H. Soifer}
\affiliation{Stanford Institute for Materials and Energy Sciences, SLAC National Accelerator Laboratory, 2575 Sand Hill Road, Menlo Park, CA 94025, USA}
\author{H. Eisaki}
\affiliation{Electronics and Photonics Research Institute, National Institute of Advanced Industrial Science and Technology, Tsukuba, Ibaraki 305-8558, Japan}
\author{P.~S. Kirchmann}
\email{kirchman@slac.stanford.edu}
\affiliation{Stanford Institute for Materials and Energy Sciences, SLAC National Accelerator Laboratory, 2575 Sand Hill Road, Menlo Park, CA 94025, USA}
\author{Z.-X. Shen}
\email{zxshen@stanford.edu}
\affiliation{Stanford Institute for Materials and Energy Sciences, SLAC National Accelerator Laboratory, 2575 Sand Hill Road, Menlo Park, CA 94025, USA}
\affiliation{Geballe Laboratory for Advanced Materials, Departments of Physics and Applied Physics, Stanford University, Stanford, CA 94305, USA}

\date{\today}

% ***** ABSTRACT
\begin{abstract}

Cuprate superconductors host a multitude of low-energy optical phonons. Using time- and angle-resolved photoemission spectroscopy, we study coherent phonons in Bi$_{2}$Sr$_{2}$Ca$_{0.92}$Y$_{0.08}$Cu$_{2}$O$_{8+\delta}$. Sub-meV modulations of the electronic band structure are observed at frequencies of $3.94\pm 0.01$ and $5.59\pm 0.06$~THz. For the dominant mode at $3.94$~THz, the amplitude of the band energy oscillation weakly increases as a function of momentum away from the node. Theoretical calculations allow identifying the observed modes as CuO$_{2}$-derived $A_{1g}$ phonons. The Bi- and Sr-derived $A_{1g}$ modes which dominate Raman spectra in the relevant frequency range are absent in our measurements. This highlights the mode-selectivity for phonons coupled to the near-Fermi-level electrons, which originate from CuO$_{2}$ planes and dictate thermodynamic properties.

\end{abstract}
\maketitle

Understanding electron-phonon coupling (EPC) has been crucial to the study of superconductivity. In conventional superconductors, EPC facilitates Cooper pair formation~\cite{Bardeen1957}. In cuprate high temperature superconductors, a complex phonon spectrum is observed~\cite{Renker1989,Reznik2006,Fong1995,Cardona1988, Sugai1989, Boekholt1990, Denisov1989, Liu1992, Kakihana1996, Sugai2003}, and the contribution of particular phonon modes to superconductivity remains debated. It is thus important to study individual phonon modes and their respective coupling to the electronic states in cuprates.

\begin{figure*}[htp]
\begin{center}
\includegraphics[width=7in]{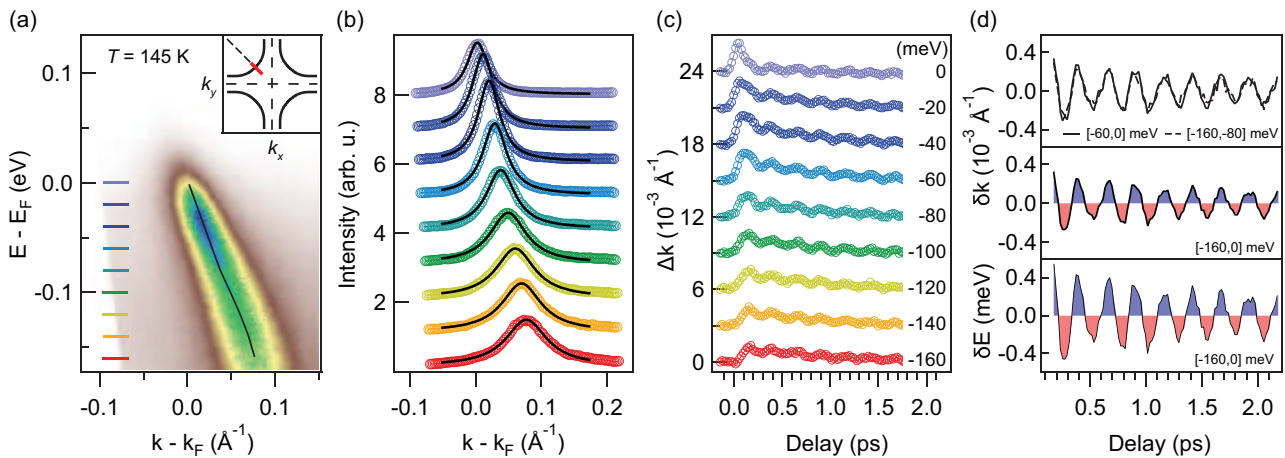}
\caption{(color online). Oscillations of the nodal band dispersion at $T = 145$~K with an incident pump fluence of $0.28$~mJ/cm$^{2}$. (a) ARPES spectrum at the node before time zero. The overlaid band dispersion (black) is obtained by fitting the momentum distribution curves (MDCs). Inset illustrates the momentum-space trajectory (red) and the Fermi surface calculated by a tight-binding model (black)~\cite{Markiewicz2005}. (b) MDCs for select binding energies from $0$ to $-160$~meV, as indicated by colored markers in panel (a). The Lorentzian fits (black) are overlaid on the MDCs. (c) Momentum dynamics ($\Delta k$) extracted from the time-dependent MDCs. Traces are offset for clarity. (d) Coherent responses extracted from the momentum dynamics. The coherent momentum oscillation ($\delta k$) is extracted by removing the smooth backgrounds in panel (c), and integrated within the energy windows of $[-60,0]$~meV (solid black), $[-160,-80]$~meV (dashed black), and $[-160,0]$~meV (blue-red shade), respectively. The coherent energy oscillation ($\delta E$) is obtained by multiplying $\delta k$ with the Fermi velocity $1.86$~eV.\AA (blue-red shade, bottom).}
\label{Fig1}
\end{center}
\end{figure*}

A variety of spectroscopic techniques have been utilized to study EPC in cuprates. Phonon spectroscopies such as Raman spectroscopy~\cite{Cardona1988, Sugai1989, Boekholt1990, Denisov1989, Liu1992, Kakihana1996, Sugai2003}, optical spectroscopy~\cite{Tsvetkov1999,Kovaleva2004}, inelastic x-ray scattering~\cite{LeTacon2013,He2018} and neutron scattering~\cite{Fong1995, Reznik2006} have been extensively applied to cuprates. However, these scattering experiments do not directly resolve the coupling to electronic states. Angle-resolved photoemission spectroscopy (ARPES) resolves disperson kinks in the electron momentum space, indicating strong coupling between electrons and bosonic modes~\cite{Gromko2003,Kaminski2001,Lanzara2001,Cuk2004,Devereaux2004,Lee2008,Iwasawa2008,Zhou2005,Meevasana2006}. These kinks have been assigned to EPC involving oxygen-derived phonons~\cite{Lanzara2001,Cuk2004,Devereaux2004,Lee2008,Iwasawa2008,Zhou2005,Meevasana2006}. However, it remains difficult to separate the contributions from individual phonon modes in the analysis of dispersion kinks~\cite{Zhou2005,Meevasana2006}.

Optical pump-probe experiments~\cite{Gadermaier2010, DalConte2012} have been employed to study EPC in cuprates based on the two-temperature model~\cite{Allen1987}. These experiments extract the EPC strength by tracking the changes of the optical reflectivity due to photoexcited electrons. However, the extracted EPC strength is averaged over the entire Fermi surface and all phonon modes, and is based on the assumption that electrons and phonons can be treated as instantaneously thermalized ensembles~\cite{Allen1987}.

Ultrafast excitation of coherent phonon oscillations simultaneously modulates the lattice and electronic properties at the same frequency. This provides an opportunity to resolve phonon frequencies with a resolution $<0.1$~THz ($0.4$~meV), thus enabling a mode-specific investigation of EPC~\cite{Sobota2014,Leuenberger2015,Gerber2017}. For cuprate superconductors, time-resolved optical spectroscopies have observed coherent phonons~\cite{Chwalek1990,Albrecht1992,Okano2011,Takahashi2009,Hinton2013,Torchinsky2013,Mansart2013,Misochko2000}, yet the inability to directly resolve electronic states limits the understanding on microscopic interactions. 

\begin{figure}
\begin{center}
\includegraphics[width=\columnwidth]{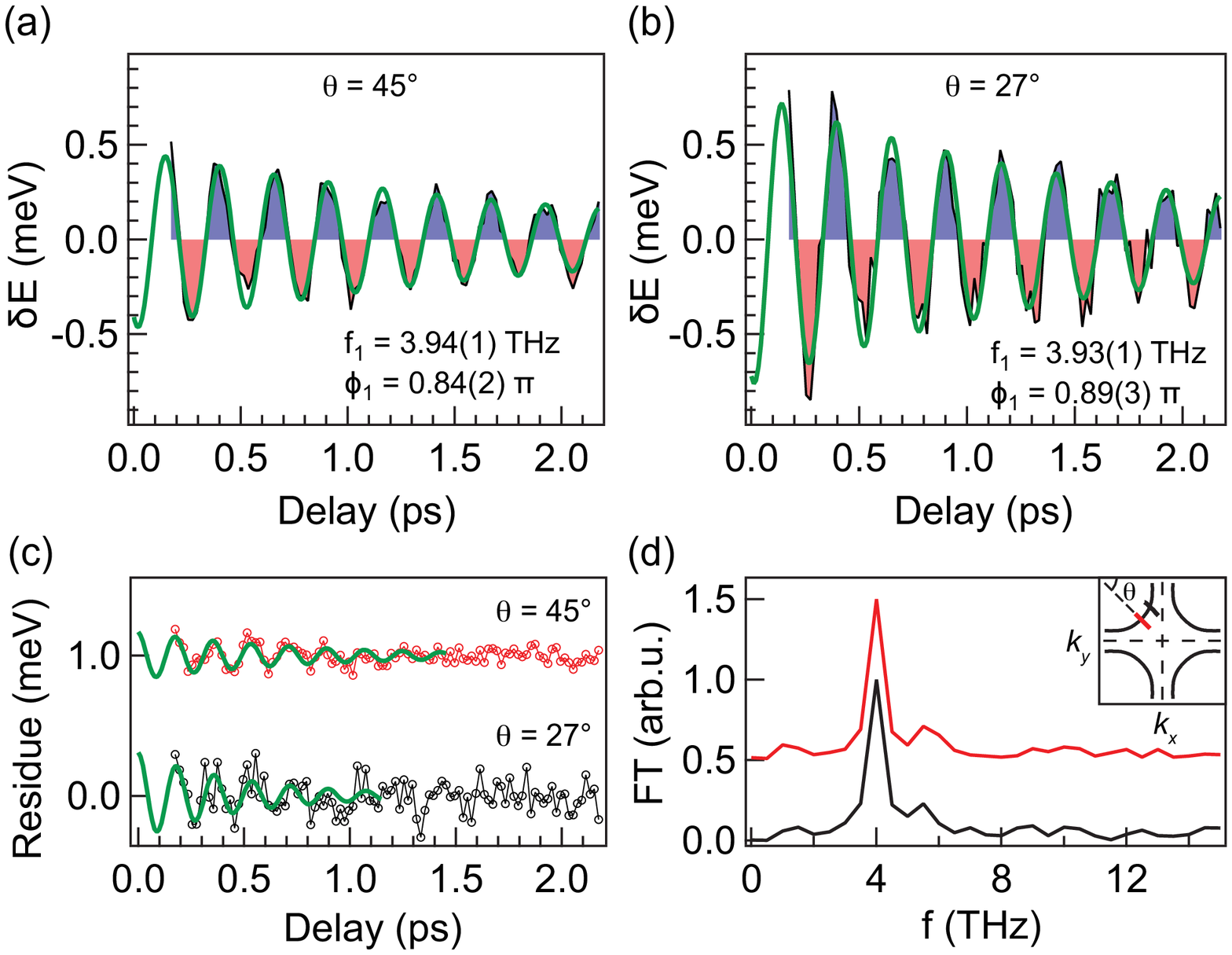}
\caption{(color online). Frequency analysis of coherent phonon modes. (a,~b) Coherent energy oscillations (blue-red shades) at (a) $\theta = 45$\textdegree~(node) and (b) $\theta = 27$\textdegree. Fitting to an exponentially-decaying cosine function (green) yields the phonon frequency $f_{1}$ and phase $\phi_{1}$. (c) Fitting residues (circles) from the analysis in (a) and (b) exhibit a distinct coherent mode. Fitting the residual data to exponentially-decaying cosine functions yields $f_{2} = 5.59\pm 0.06$~THz with a phase $\phi_{2} = 0.05\pm 0.08$~$\pi$. (d) Fourier transforms (FTs) of the coherent energy oscillations for $\theta = 45$\textdegree~(red) and $27$\textdegree~(black). The momentum trajectories are indicated in the inset.}
\label{Fig2}
\end{center}
\end{figure}

\begin{figure}
\begin{center}
\includegraphics[width=\columnwidth]{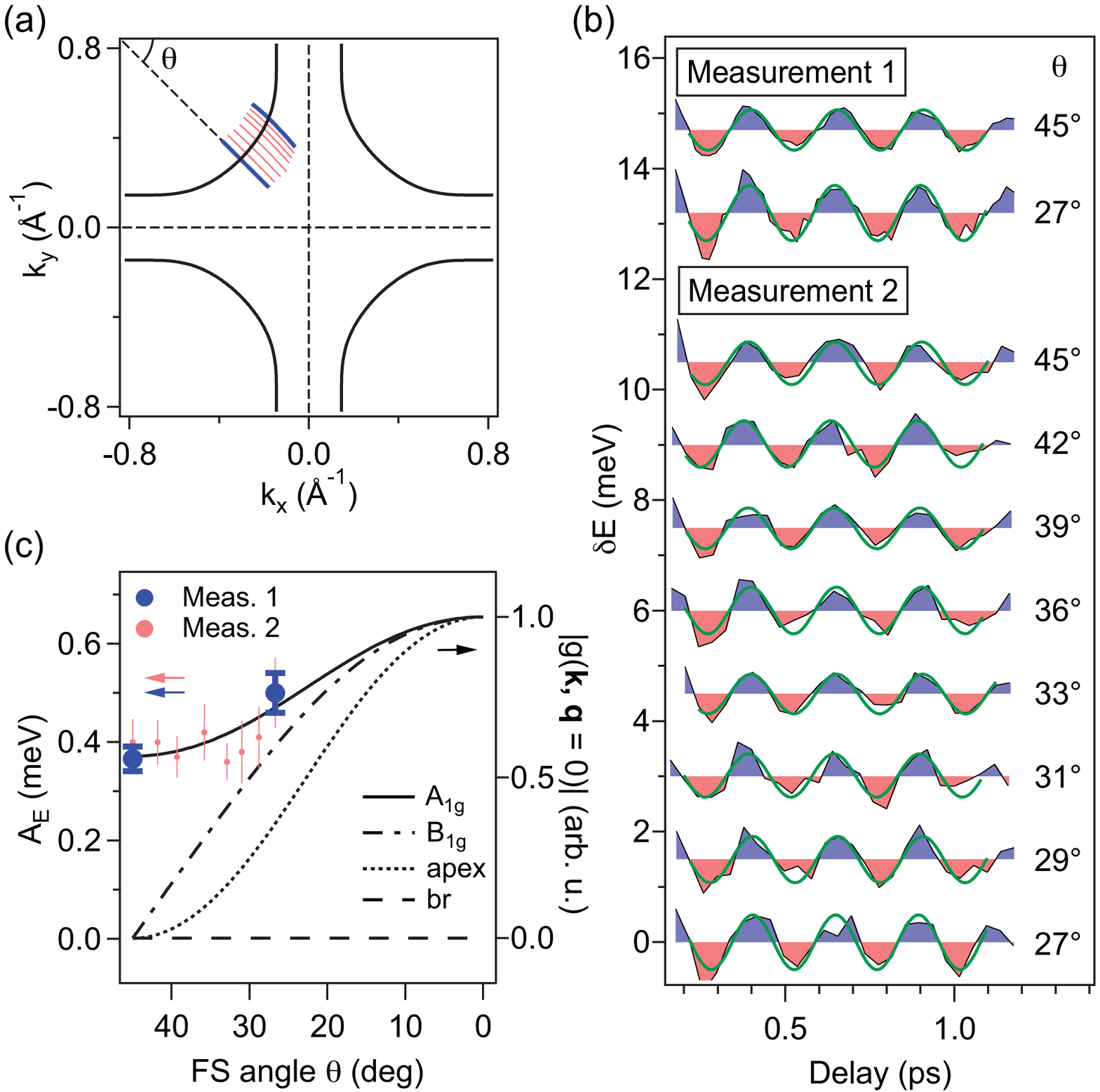}
\caption{(color online). Momentum dependence of the coherent phonon oscillation. (a) Fermi surface (FS) plot (black) with contours (red and blue) indicating the momenta for trARPES measurements. (b) Coherent energy oscillations as a function of the FS angle. The absorbed energy density of $22$~meV/(pulse$\cdot$unit cell) is enforced while rotating the sample~\cite{SOM}. (c) Amplitudes of the energy oscillations as a function of the FS angle. Measurement~1 is emphasized by thicker markers considering its higher signal-to-noise and smaller fitting uncertainties. Overlaid are the theoretical EPC strengths (lines) for the $A_{1g}$ and $B_{1g}$ modes, the apical mode (apex), and the breathing mode (br), respectively. The theoretical results are normalized by their respective maxima in the Brillouin zone.}
\label{Fig3}
\end{center}
\end{figure}

Notably, the EPC strength is characterized by the momentum-dependent deformation potential $D({\bf k})$, which is defined by the ratio of the electronic energy shift $\delta\epsilon ({\bf k})$ and the corresponding lattice distortion $\delta u$: $D({\bf k}) = \delta\epsilon({\bf k})/\delta u$~\cite{Khan1984}. Time-resolved ARPES (trARPES) is an ideal tool to access $\delta\epsilon({\bf k})$ as the lattice vibrates. Among others, this technique has been applied to Bismuth~\cite{Papalazarou2012,Greif2016}, Bi$_{2}$Se$_{3}$~\cite{Sobota2014}, Bi$_{2}$Te$_{3}$~\cite{Golias2016}, CeTe$_{3}$~\cite{Leuenberger2015}, and FeSe/SrTiO$_{3}$ thin films~\cite{Gerber2017}, yielding important insight on the momentum-dependent EPC in these materials. Meanwhile, despite the significant progress in revealing quasiparticle dynamics in cuprates using trARPES~\cite{Perfetti2007,Graf2011,Cortes2011,Smallwood2012,Zhang2014,Yang2015a,Miller2017}, there has not been any trARPES study reporting coherent phonons in cuprates.

In this Letter we report the first trARPES study of coherent phonon oscillations in a cuprate superconductor. In optimally doped Bi$_{2}$Sr$_{2}$Ca$_{0.92}$Y$_{0.08}$Cu$_{2}$O$_{8+\delta}$ (OP Bi2212, T$_{c}$ = $96$~K), we discover a dominant modulation of the electronic band structure at $3.94\pm 0.01$~THz, and a sub-dominant modulation at $5.59\pm 0.06$~THz. Our study reveals that the oscillation amplitude of the dominant mode weakly increases from the node to the antinode, which reflects the momentum dependence of the deformation potential for an $A_{1g}$ phonon involving CuO$_{2}$ motions~\cite{Johnston2010, Devereaux2016}. The ability to precisely determine the phonon frequency permits us to discern that the electronic states near the Fermi level $E_{\rm F}$ are selectively coupled to the CuO$_{2}$ related $A_{1g}$ phonon, rather than the Bi and Sr related phonons which dominate Raman spectra in the relevant frequency range~\cite{Cardona1988, Sugai1989, Boekholt1990, Denisov1989, Liu1992, Kakihana1996, Sugai2003}. This illustrates the mode- and band-selective nature of EPC in complex materials such as cuprates and the power of trARPES to uniquely address it. 

Our trARPES system is based on a Ti-Sapphire regenerative amplifier which outputs $1.5$~eV, $35$~fs pulses at a repetition rate of $312$~kHz. The fundamental beam is split into pump and probe paths. In the pump path, a mechanical translation stage tunes the pump-probe delay. In the probe path, two stages of second harmonic generation provide $6$~eV probe pulses. Pump and probe beam profiles are characterized by the full-widths-at-half-maximum (FWHMs), which are $154\times 166$~$\mu$m$^{2}$ and $50\times 103$~$\mu$m$^{2}$, respectively. The overall time resolution is $77$~fs as characterized by a resolution-limited cross-correlation~\cite{SOM}. We use an incident pump fluence of $0.28$~mJ/cm$^{2}$. Photoelectrons are collected by a Scienta R4000 hemispherical analyzer, under ultrahigh vacuum with a pressure $<7\times 10^{-11}$~Torr. The overall energy resolution is $40$~meV. The measurement temperature is $145$~K. 

The band dispersion along the Brillouin zone diagonal (node) before time zero is displayed in Fig.~\ref{Fig1}(a). We extract band dispersions by fitting the momentum distribution curves (MDCs) to Lorentzian functions (Fig.~\ref{Fig1}(b)). The extracted band dispersion at $145$~K does not exhibit strong dispersion kinks, which is likely due to both the $40$~meV energy resolution and thermal spectral broadening~\cite{Grimvall1981}. Figure~\ref{Fig1}(c) displays the change of MDC peak positions ($\Delta k$) as a function of pump-probe delay for select binding energies. We identify periodic oscillations in the momentum dynamics on top of incoherent dynamics. 

To investigate the binding energy dependence of the coherent response, we define two energy integration windows with respect to $E_{\rm F}$: $[-160,-80]$ and $[-60,0]$~meV. We remove the incoherent signal by fitting the momentum dynamics to a $4^{\rm th}$-order polynomial function and extracting the fitting residuals $\delta k$~\cite{Yang2015}. In the top panel of Fig.~\ref{Fig1}(d) we compare $\delta k$ integrated within the two energy windows. The oscillation amplitude is independent of binding energy within experimental uncertainties. Therefore we integrate the coherent response over the entire energy range of $[-160,0]$~meV for an improved signal-to-noise ratio (SNR), as shown in the middle panel of Fig.~\ref{Fig1}(d). Since the band dispersion is approximately linear (Fig.~\ref{Fig1}(a)), we cannot distinguish momentum shifts versus energy shifts. We calculate band energy oscillations via $\delta E = v_{\rm F}\delta k$, where $v_{\rm F} = 1.86$~eV$\cdot$\AA~is the nodal Fermi velocity (Fig.~\ref{Fig1}(d), bottom panel).

The coherent energy oscillations $\delta E$ at the node and $18$\textdegree~off the node are obtained using their corresponding Fermi velocities, and presented in Fig.~\ref{Fig2}(a) and (b). These oscillations are fitted to an exponentially-decaying cosine function, 

\begin{equation}\label{eqn1}
A\exp{(-t/\tau)}\cos{(2\pi f t+\phi)}
\end{equation}

which yields the frequency $f$ and phase $\phi$ for each data set. $A$ and $\tau$ stand for the oscillation amplitude and the relaxation time constant, respectively. $t$ is the time delay, with time zero independently determined by a resolution-limited cross-correlation~\cite{SOM}. The extracted frequency and phase of the dominant mode are $f_{1} = 3.94\pm 0.01$~THz and $\phi_{1} = 0.84\pm 0.02$~$\pi$, respectively, and are consistent across nodal and off-nodal measurements within fitting uncertainties. The fitting residues in Fig.~\ref{Fig2}(c) display weak but discernible oscillations. This is particularly evident in the residual data at $\theta = 45$\textdegree, for which fitting using Eqn.~\ref{eqn1} yields a second mode at $f_{2} = 5.59\pm 0.06$~THz with a phase $\phi_{2} = 0.05\pm 0.08$~$\pi$. In Fig.~\ref{Fig2}(d) we plot the normalized Fourier transforms (FTs) corresponding to the nodal and off-nodal data. Both FTs confirm the existence of a dominant modulation near $4$~THz and a sub-dominant modulation near $5.6$~THz. The phase difference between these two modulations is close to $\pi$.

We present a detailed momentum-dependent study of the dominant $3.94$~THz mode in Fig.~\ref{Fig3}. To compare results obtained with different sample orientations, we enforce a constant absorbed energy per pulse per unit cell while rotating the sample ($22$~meV/(pulse$\cdot$unit cell))~\cite{Leuenberger2015,SOM}. We also consider the anisotropy of optical constants and the effect of time-resolution broadening due to the change in incident angles. These factors only contribute to $<5\%$ uncertainties of the oscillation amplitudes~\cite{SOM}. Results from two measurements on the same sample are displayed in Fig.~\ref{Fig3}(b). Measurement 1 refers to the same data as shown in Fig.~\ref{Fig2}, which covers only the two extreme Fermi surface (FS) angles with a high SNR. In contrast, Measurement 2 examines 8 FS angles with a lower SNR. Sub-meV coherent energy oscillations are resolved for all FS angles and both measurements. Due to the limited SNR in Measurement 2, we fit the oscillations to non-decaying cosine functions to extract average amplitudes in the time window of $[0.2, 1.1]$~ps. The extracted average amplitude $A_{E}$ shows a weak momentum dependence (Fig.~\ref{Fig3}(c)). 

The momentum-resolved oscillation amplitude directly reflects the underlying deformation potential $D({\bf k})$ for ${\bf q} = 0$ modes~\cite{Kuznetsov1994, Khan1984}. $D({\bf k})$ is proportional to the electron-phonon coupling vertex $g({\bf k,q=0})$ which is more commonly discussed in theoretical studies~\cite{Khan1984,Grimvall1981}.

\begin{equation}\label{eqn2}
g({\bf k, q=0}) = \left( \frac{\hbar}{2MN\omega} \right)^{1/2} \hat{\boldsymbol \epsilon} \cdot {\bf D(k)} \propto A_{E}({\bf k})
\end{equation}

Here $\hbar$, $M$, $N$, $\omega$, and $\hat{\boldsymbol \epsilon}$ represent the reduced Planck constant, the mode effective mass, the number of unit cells, the phonon frequency, and the phonon eigenvector, respectively. Eqn.~\ref{eqn2} allows us to use the experimental oscillation amplitude $A_{E}$ to compare to theoretical predictions of $g({\bf k, q = 0})$.

Johnston {\it et al.} provides a comprehensive theoretical survey of several important oxygen phonons in cuprates based on the charge-transfer induced coupling~\cite{Johnston2010}. The formalism can be significantly simplified for ${\bf q = 0}$ modes which are coherently driven by optical excitations. Here we summarize the EPC vertices for the in-plane breathing mode (``br"), the out-of-plane A$_{1g}$ and B$_{1g}$ modes, and the apical oxygen mode (``apex").

\begin{eqnarray}\label{eqn3}
g_{\rm br}({\bf k, q=0}) &=& 0\nonumber\\
g_{\rm A_{1g}, B_{1g}}({\bf k, q=0}) &\sim&  \sin^{2}{(k_{x}a/2)} \pm \sin^{2}{(k_{y}a/2)}\nonumber\\
g_{\rm apex}({\bf k, q=0}) &\sim& \left[ \cos{(k_{x}a/2)} - \cos{(k_{y}a/2)} \right]^{2}
\end{eqnarray}

Here $a$ is the lattice constant assuming a tetragonal unit cell. In Fig.~\ref{Fig3}(c) we plot the EPC vertices for various phonon modes using Eqn.~\ref{eqn3}. The theoretical EPC vertices are normalized by their respective maxima in the Brillouin zone. Notably, only $A_{1g}$ modes exhibit non-zero coupling at the node and their coupling strength increases slightly as the momentum moves away from the node. In addition, the coherent response in time-resolved spectroscopies is generally dominated by $A_{1g}$ modes which inherit the full lattice symmetry~\cite{Gerber2015,Gerber2017,Sobota2014,Li2013}. We thus conclude that the $3.94$~THz modulation corresponds to an $A_{1g}$ mode in the tetragonal notation, or an $A_{g}$ mode in the orthorhombic notation~\cite{Kakihana1996,Sugai1989,Sugai2003}. Similarly, the $5.59$~THz mode is likely a fully symmetric mode considering its nonzero coupling near the node.

We discuss the significance of our observations in comparison to the previous literature on optical phonons in Bi2212. We focus on low-energy modes in the frequency range of $0\sim 6$~THz ($0\sim 200$~cm$^{-1}$), as the amplitudes of coherent oscillations at higher frequencies are significantly reduced due to the finite time resolution~\cite{Gerber2017}. Bi2212 has 6 $A_{1g}$ modes based on the $I4/mmm$ tetragonal structure~\cite{Liu1992,Kakihana1996,Kovaleva2004}. Near optimal doping, the pronounced $A_{1g}$ modes in Raman spectra are at $1.8$ and $3.6$~THz ($60$ and $120$~cm$^{-1}$)~\cite{Cardona1988, Sugai1989, Boekholt1990, Denisov1989, Liu1992, Kakihana1996, Sugai2003}. A time-resolved reflectivity experiment also observes these two modes, verifying that ultrafast optical excitations at $1.5$~eV can indeed launch these modes coherently~\cite{Misochko2000}. Meanwhile, in our trARPES experiment the $1.8$ and $3.6$~THz modes are below the noise level in the Fourier transform of the coherent response (Fig.~\ref{Fig2}(d)). The dominant mode in trARPES is instead at $3.94$~THz ($\sim 130$~cm$^{-1}$), and we observe it for a wide range of dopings~\cite{SOM}. This mode only appears as a weak shoulder-like feature in Raman spectra~\cite{Kakihana1996, Sugai2003}. These discrepancies challenge our understanding of the coupling between electrons and low-energy optical phonons in Bi2212.

We resolve these discrepancies by considering the nature of different phonon modes. From Raman spectroscopy, the $1.8$ and $3.6$~THz modes have been assigned to the $A_{1g}$ phonons mainly involving Bi and Sr motions~\cite{Sugai1989, Boekholt1990, Denisov1989, Liu1992, Kakihana1996, Falter2003, Kovaleva2004}. These modes are unlikely to be strongly coupled to the electrons near $E_{\rm F}$ due to the spatial separation between the CuO$_{2}$-derived electrons and the atomic motions in the charge reservoir layers. On the other hand, the assignment of the $3.94$~THz mode in the Raman literature has been debated. While some studies assign it to a Cu $A_{g}$ mode in the orthorhombic notation~\cite{Sugai1989,Liu1992}, others attribute it to disorder-induced Sr- or Bi-derived vibrations~\cite{Kakihana1996}. Based on the momentum dependence of the coupling strength (Fig.~\ref{Fig3}(c)), our study supports the assignment to a Cu $A_{g}$ mode. This is also consistent with the theoretical framework in Ref.~\cite{Johnston2010}. These considerations highlight an important fact: Raman spectroscopy and time-resolved reflectivity measurements are powerful tools for identifying phonon modes, but do not directly reflect how the modes couple to the electronic bands near $E_{\rm F}$. Moreover, a time-resolved reflectivity experiment on La$_{1.85}$Sr$_{0.15}$CuO$_{4}$ shows that different probe photon energies reveal coherent modes at different frequencies, emphasizing the nontrivial correspondence between electronic bands and the phonon modes they couple to~\cite{Mansart2013}. trARPES probes the EPC through the band-specific low-energy electron dynamics, and allows us to identify which of the many phonons in cuprates are coupled to the electronic band responsible for transport and thermodynamic properties.

%On the other hand, the $3.94$~THz mode is associated with atomic motions in the CuO$_{2}$ planes, allowing a direct coupling to the electrons near $E_{\rm F}$~\cite{Sugai1989, Sugai2003}.

Our experiment is complementary to equilibrium ARPES, which probes EPC via the analysis of dispersion kinks~\cite{Gromko2003,Kaminski2001,Lanzara2001,Cuk2004,Devereaux2004,Lee2008,Iwasawa2008,Zhou2005,Meevasana2006,Vishik2010,Kondo2013}. In particular, high-resolution ARPES studies have revealed EPC involving multiple phonon modes in the range of $2.4\sim 7.3$~THz ($10\sim 30$~meV)~\cite{Vishik2010,Kondo2013}. Meanwhile, it is difficult to quantify the coupling to individual modes due to the complication of bare-band dispersions and the effective integration over all phonon momenta. In contrast, coherent phonon studies using trARPES avoid the complication of bare-band dispersions, and are only sensitive to ${\bf q = 0}$ modes. Moreover, time-domain measurements provide higher sensitivity to lower-frequency optical modes. The effective phonon energy resolution is only limited by the time range of coherent oscillations. Therefore, different modes which are closely spaced in energy can be distinguished in the time domain and selectively investigated~\cite{Sobota2014}.

In this study we measure the momentum dependence of EPC strength in cuprate superconductors, enabled by the capabilities of trARPES to resolve electronic bands modulated by phonon modes. As a step beyond traditional phonon spectroscopies, our measurement identifies two $A_{1g}$ modes which selectively couple to the electronic states near $E_{\rm F}$. An immediate extension of the present study is to combine with a time-resolved diffraction measurement, which tracks the lattice distortion $\delta u$ for the corresponding phonon oscillation. Combining the electron energy shift and the lattice distortion defines a coherent ``lock-in" experiment which determines the deformation potential purely experimentally~\cite{Gerber2017}. Furthermore, improvements of the time resolution will grant access to other important modes, in particular the $8.5$~THz $B_{1g}$ mode~\cite{Devereaux2004,Cuk2004,Lee2008,He2018a} and the $17$~THz apical mode~\cite{Pashkin2010}. These experimental pursuits will be vital for theories examining the complex interactions underlying high temperature superconductivity.

\begin{acknowledgments}
{\bf Acknowledgments} We thank Brian Moritz and Thomas Devereaux for fruitful discussions. This work was supported by the U.S. Department of Energy, Office of Science, Basic Energy Sciences, Materials Sciences and Engineering Division under contract DE-AC02-76SF00515. S.-L.Y. acknowledges support by the Stanford Graduate Fellowship. S.-L.Y. is also supported by the Kavli Postdoctoral Fellowship at Cornell University. J.A.S. is in part supported by the Gordon and Betty Moore Foundations EPiQS Initiative through Grant GBMF4546. D.L. acknowledges partial support by the Swiss National Science Foundation under fellowship P300P2151328. H.S. acknowledges support from the Fulbright Scholar Program.
\end{acknowledgments}

\bibliography{YangSL_Bi2212_CoPho_refs}

\end{document}